# Concave Plasmonic Particles: Broad-Band Geometrical Tunability in the Near Infra-Red

*Nikolai Berkovitch\*, Pavel Ginzburg, and Meir Orenstein.*

Department of Electrical Engineering, Technion, Haifa 32000, Israel.

\*E-mail: nikolaib@tx.technion.ac.il

Optical resonances spanning the Near and Short Infra-Red spectral regime were exhibited experimentally by arrays of plasmonic nano-particles with concave cross-section. The concavity of the particle was shown to be the key ingredient for enabling the broad band tunability of the resonance frequency, even for particles with dimensional aspect ratios of order unity. The atypical flexibility of setting the resonance wavelength is shown to stem from a unique interplay of local geometry with surface charge distributions.

Metallic nanoparticles have been widely studied due to their interesting optical properties associated with localized plasmon resonances. These resonances have a variety of applications including: enhanced sensing and spectroscopy[1], plasmonic biosensors[2], cancer imaging and therapy[3,4], building block of metamaterials[5,6], ability to redirect scattered light[7], plasmonic lasers[8], SPASERS[9], enhanced nonlinearities[10], enhancement of radiation efficiencies[11,12], and much more. Typical metallic nanoparticle's geometries, such as spheres, disks[13], cups[7], and bow-ties[14,15] on a quartz or glass substrate, exhibit resonances primarily in the 'visible' spectrum[16,17] (below 1 micrometer). The extension of these resonances to the near and short infra red (IR) regime (1-2µm) is of great importance for optical communications, bio-medical applications and much more. For these purposes extreme modifications to the particles' geometry should be applied, e.g.: coupled particles separated by only several nanometers[17,18,19], few-nanometers-thin nano-rings[3], and elongated particles with very large aspect ratios[20]. Thus the repeatable fabrication of such configurations is challenging and limiting their applicability. High aspect ratio (elongated) particles are also difficult to integrate as metamaterials' unit cells, since unit cell dimension should be substantially smaller than the light wavelength.

Here we present metamaterials based on concave plasmonic unit cells. The specific unit cells are comprised of nano-scale cylinders with cross sections varied between the more conventional convex to highly concave shapes. The resonances of such metamaterials are significantly modified by the convexity sign of the cylinder cross section. Large tunability (many hundreds of nanometers) is achieved by slightly tuning geometrical parameters of concave particles, which is feasible without either pushing the fabrication process to extremes (compared e.g. to the few-nanometer-thin rings required for similar tunability range) or largely distorting the square unit cell proportions (such distortion enhances considerably the particle size which yields higher order excitations as well as in inter-cell coupling). We detail the experimental and numerical results and provide the conceptual interpretation of the resonances. Specifically, we show that concave particles with dimensional aspect ratios of ~ 1 can support selectable resonance wavelengths encompassing the whole near and short IR regime.

The metamaterial is comprised of gold nanoparticles' arrays, which were produced by electron beam lithography (EBL) on a glass substrate covered with 30nm ITO layer to eliminate charging effects. The nanoparticles are separated by gaps of about 300nm in order to avoid coupling between adjacent unit cells. The thickness of the nanoparticles is 60nm and their lateral dimension is varied in the 100 nanometers regime. The measurement of these metamaterials was accomplished by illuminating them with polarized white light under normal incidence and by monitoring the spectral features of the transmitted light (Fig.1).

The localized plasmon resonance of a nanodisk is dependent on its geometry and dielectric properties of the disk and surrounding media. The resonance conditions of regular convex nanodisk can be analytically approximated as those of an ellipsoid with major axes $a,b,c$ under Mie scattering[21]:

$$\alpha_i = V \cdot \frac{\varepsilon - \varepsilon_m}{\varepsilon_m + L_i(\varepsilon - \varepsilon_m)}, \quad i = a,b,c \tag{1}$$

where $\varepsilon$ and $\varepsilon_m$ are dielectric constants of the particle and the surrounding medium respectively, $V$ the particle's volume, $L_i$ the geometrical factor along the ellipsoid axis $i$, and $\alpha_i$ the polarizability tensor. In particular, we are interested in the normal incidence excitation of a fixed thickness $(h = 60nm)$ disk, starting with lateral diameter of 100nm (on a glass substrate with $n = 1.5$). It is evident, that distorting the disk cross section by increasing the diameter parallel to the electrical field direction $\vec{E}$ results in a red shift of the resonance, while enhancing the disk diameter in a perpendicular direction yields a small blue shift (Fig.2). In order to shift the basic plasmon resonance of a 100nm diameter Au sphere, located deep in the visible part of spectrum (~540nm), to the center of the NIR regime ($\lambda = 1550nm$), we should distort sphere diameter in the field direction to $420nm$ (aspect ratio of 4.2). Such a large ratio yields excitation of additional higher order modes (also retarded modes), as well as strongly couples between the metamaterial unit cells. Thus a simple distortion of a nanodisk cannot serve for affordable extreme tunability in IR and another approach is requested.

We propose here to employ the particle's concavity parameter to achieve large scale shifting and tuning of the plasmon resonance. In order to demonstrate the idea we fabricated metamaterials comprised of particles with a cross section of different convexities (and convexity signs). All particles have equal thickness (60nm) and similar lateral dimensions: central-width W=100nm, and length L=250nm; a concave (hyperbolic) particle has an additional parameter – base-width B (inset of Fig.1b for definition). The measured transmission spectrum with field polarized along the particle length is shown in Fig.3. A metamaterial with unit cell based on a particle with rectangular cross section (aspect ratio 2.5) exhibited a resonance at 1080nm. A convex cylinder unit cell (elliptical cross section with the same aspect ratio) had a slightly blue shifted resonance (1020nm), while even a slightly concave (hyperbolic) cylinder (B=135nm) exhibited a significant red shifted resonance (1180nm).

For further investigation of the resonance tuning by the particle's concavity, we measured the resonance of concave hyperbolic particles as a function of the base-width B while keeping constant their central-width W and length L. The experimental results as well as numerical results by Finite Difference Time Domain (FDTD) are depicted in Fig.4. The resonance is considerably (more than 350nm) red-shifted by increasing the base-width of the concave particle (from 130 to 300nm). This result has an opposite trend compared to that obtained for convex disk particles, where enhancing the size of the particle perpendicular to the field polarization results in a small blue shift (Fig.2). Thus the red-shift encountered here is indeed related to the degree of concavity of the unit cell and not to the increasing the effective dimension (or area) of the particle.

The fundamentally different resonance behavior of convex and concave particles is related to the combination of global geometrical shaping of the surface charge distribution and the local geometry at points of high surface charge density. In the electro-quasi-static regime the geometry dependent eigen-solutions for the surface charge density and the eigen-values, related to the resonance frequency, are given by the following Fredholm integral equation[22]:

$$\sigma(Q) = \frac{\varepsilon(\omega)-1}{\varepsilon(\omega)+1} \oint_S \sigma(M) \frac{\vec{r}_{MQ} \cdot \hat{n}_Q}{\pi |\vec{r}_{MQ}|^2} \cdot dS_M \qquad (2)$$

where $\sigma(Q)$ is a surface charge density at point Q, $\varepsilon(\omega)$ the particle dielectric constant, $\vec{r}_{MQ}$ is a vector, connecting two points on the particle boundaries: any point (M) with a point of observation (Q), $\hat{n}_Q$ is a normal to the boundary at the point Q, and the integration is performed on the particle boundary. Eq. 2 here is two-dimensional (for simplicity), while a full-fledged three-dimensional equation is written as a surface integral. It may be shown[22], that the resonance frequency for convex particles (as determined by the eigen-value $\lambda = [\varepsilon(\omega)-1]/[\varepsilon(\omega)+1]$) is bounded from below by geometrical factors, which necessitates very large aspect ratios in order to achieve IR resonances. At these aspect ratios and for affordable fabrication parameters higher order resonances are invoked, and, furthermore, the quasi-static approximation may be invalid as well. However, this fundamental limit is not encountered for concave geometry, causing the substantial difference in resonance characteristics.

For the exact solution of the resonance frequency, Eq. 2 (or its 3D version) can be solved numerically. However the conceptual explanation for the concave particles resonance can be extracted from the structure of the integral of Eq. 2. For concave particles, the normal to the surface can point in a counter direction to the charge separation vector, resulting in a local negative value of the scalar product within the integrand of Eq. 2, which yields a reduced value of the resonance frequency whenever a significant surface charge distribution is generated on the concave edges. This is exemplified schematically in Figs. 5(a)-5(c). The surface charge density is proportional to the discontinuity in the normal component of electrical field. In Fig.5 (a) (convex particle), the two well defined groups of surface polarization charges are separated by vector $\vec{r}_{MQ}$, which is parallel to the surface normal ($\hat{n}_Q$), yielding a maximal value of $\vec{r}_{MQ} \cdot \hat{n}_Q$. This is directly related to a high resonant frequency, namely, "blue" shifted (the eigen-value $\lambda$ of Eq. 2, relying on Drude model for $\varepsilon(\omega)$, approaches unity from above, when the frequency moves towards IR). For the rectangular case (Fig. 5(b)) the surface charges are distributed along the edge such that the contributions to integral (Eq. 2) incorporate also terms with reduced magnitudes due to larger radius-vectors $\vec{r}_{MQ}$ misaligned with the surface normal $\hat{n}_Q$. Thus, the resulting integral has a lower value yielding a relative "red" shift. The most dramatic effect occurs for concave particles (Fig. 5(c)), where a significant part of the charge distribution is on the concave edges, resulting in negative scalar product in the integrand of Eq. 2 leading to a substantial reduction of the resonance frequency even for aspect ratios of the order of unity.

In conclusion, we showed experimental, numerical, and theoretical evidence, that concave nanoparticles with reasonably square-like aspect ratios exhibit broadly tunable resonances in the near and short IR. Design of particles, optimized for IR resonances, relies on generation of substantial surface charge densities on the concave edges of the particle. These easily achievable IR resonances are favorable for a large span of applications.

[1] Nie, S.; Emory, S. R., Science 1997, 275, (5303), 1102-1106.

[2] Kabashin, A. V.; Evans, P.; Pastkovsky, S.; Hendren, W.; Wurtz, G. A.; Atkinson, R.; Pollard, R.; Podolskiy, V. A.; Zayats, A. V., Nature Materials 2009,8, 867.

[3] Loo, C.; Lowery, A.; Halas, N.; West, J.; Drezek, R. Nano Lett. 2005, 5 (4) 709– 711.

[4] Loo, C.; Lin, A.; Hirsch, L.; Lee, M. H.; Barton, J.; Halas, N.; West, J.; Drezek, R. Technol. Cancer Res. Treat. 2004, 3, 33–40.

[5] Ziolkowski, R. W.; Engheta, N., Metamaterials: Physics and Engineering Explorations, IEEE Press, John Wiley & Sons, Inc., June, 2006, Chapter 1, 5 - 41.

[6] Pollard, R. J.; Murphy, A.; Hendren, W. R.; Evans, P. R.; Atkinson, R.; Wurtz, G. A.; Zayats, A. V.; Podolskiy, V. A., Phys. Rev. Lett. 2009,102, 127405.

[7] Mirin N.; Halas, N.J. 2009, Nano Lett., 9 , 1255.

[8] Hill, M. T. et al.  Nature Photon. 2007, 1, 589–594.

[9] Stockman, M. I. Nature Photonics 2008, **2**, 327-329.

[10] Wurtz, G. A.; Zayats, A. V. Laser and Photon. Rev. 2008, 2, 125-135.

[11] Khurgin, J. B.; Sun, G.; Soref, R. A. J. Opt. Soc. Am. B 2007, 24, 1968-1980.

[12] Yablonovitch, E.; Gontijo, I.; Boroditsky, M.; Keller, S.; Mishra, U.; DenBaars ,S. Physical Review B 1999, vol. 60, no.16, pp. 11564-67.

[13] Liu, Z.; Boltasseva, A.; Pedersen, R.H.; Bakker, R.; Kildishev, A.V.; Drachev, V.P.; Shalaev, V.M.; Metamaterials 2 2008, 45.

[14]  Fischer, H.;  Martin Olivier J. F. Opt. Express 2008,16, 9144-9154 .

[15] Fromm, D. P.; Sundaramurthy, A.; Schuck, P. J.; Kino, G.;Moerner, W. E. Nano Lett. 2004, 4, 957-961.

[16]  Rechberger, W.; Hohenau, A. ; Leitner, A.; Krenn, J.R.; Lamprecht, B.; Aussenegg, F.R.,  Opt. Commun.  2003, 220, 137.

[17] Atay, T.; Song, J.-H. ; Nurmikko, A. V. Nano Lett. 2004, 4, 1627–1631.

[18] Nordlander, P.; Oubre, C.; Prodan, E.; Li, K.; Stockman, M. NanoLett. 2004, 4, 899-903.

[19] Prodan, E.; Radloff, C.; Halas, N. J.; Norlander, P. Science 2003,302, 419-422.

[20] Khlebtsov, B. N.; Khlebtsov, N. G., J. Phys. Chem. C 2007, 111, 11516.

[21] Bohren, C.; Huffmann, D. Absorption and scattering of light by small particles; John Wiley & Sons: New York, 1983.

[22] Mayergoyz, I. D.; Fredkin, D. R.;  Zhang, Z., Phys. Rev. B 2005, 72, 155412 .

**FIGURE CAPTIONS**

**Figure 1**. (a) Experimental setup (b) SEM image of concave nanoparticles based metamaterial; inset shows the scheme of the concave hyperbolic particle

**Figure 2**. Influence of the geometrical factor on Mie resonance of the disk. Red solid line – increasing the disk diameter $D$ along the excitation polarization; blue dashed line – increasing the disk diameter $D$ perpendicular to the excitation polarization.

**Figure 3**. Measured transmission spectra of disk (red solid line), rectangle (blue dashed line), and concave hyperbolic particle (black dot line)

**Figure 4**. Resonance wavelength of metamaterials based on concave hyperbolic particles plotted as a function of its base-width. Measured spectra (blue rings) and FDTD result (red solid line).

**Figure 5**. Calculated (FDTD) field distributions for particle with a cross section of (a) ellipsoid (b) rectangular (c) concave. Arrows and symbols are related to Eq.2.

# Figure 1

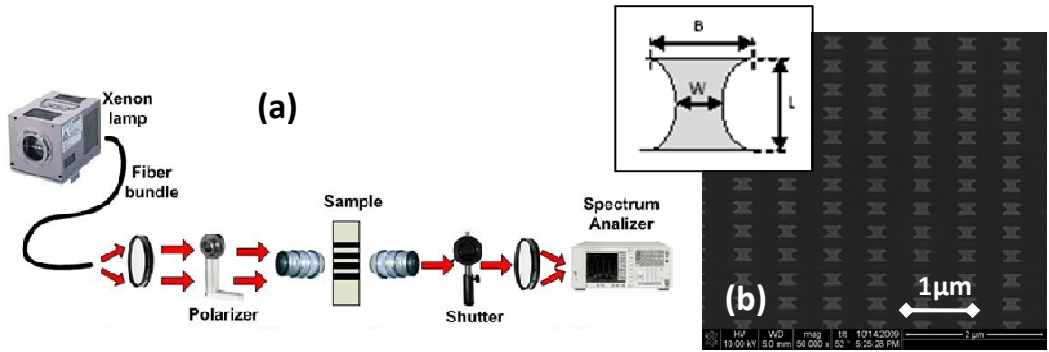

# Figure 2

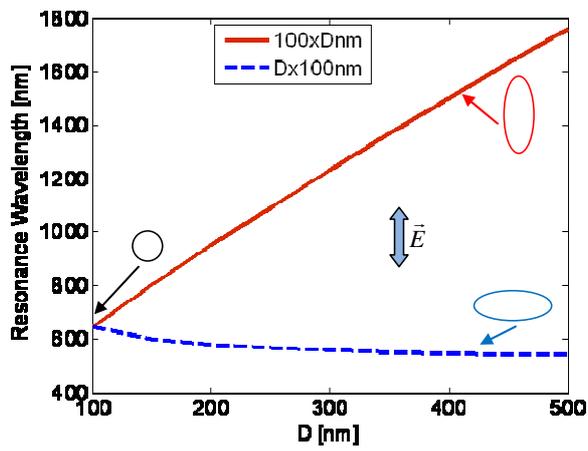

# Figure 3

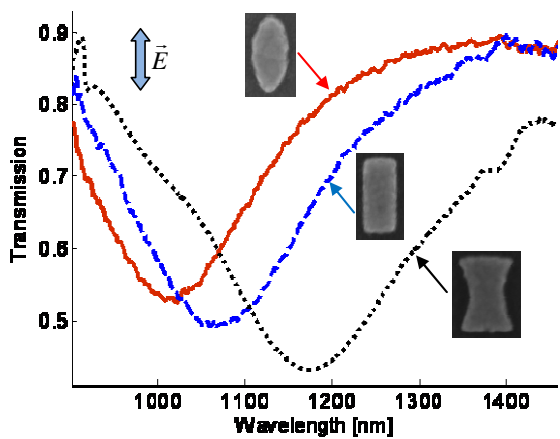

# Figure 4

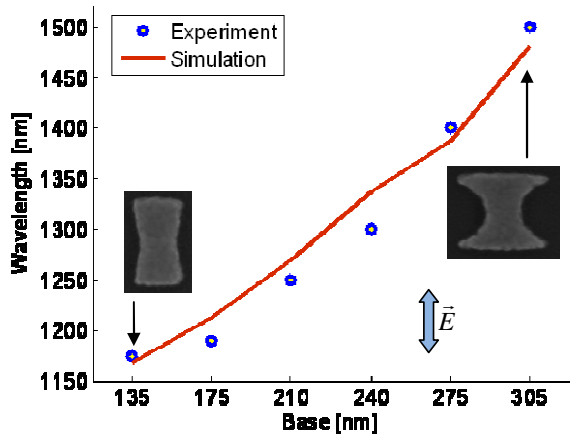

# Figure 5

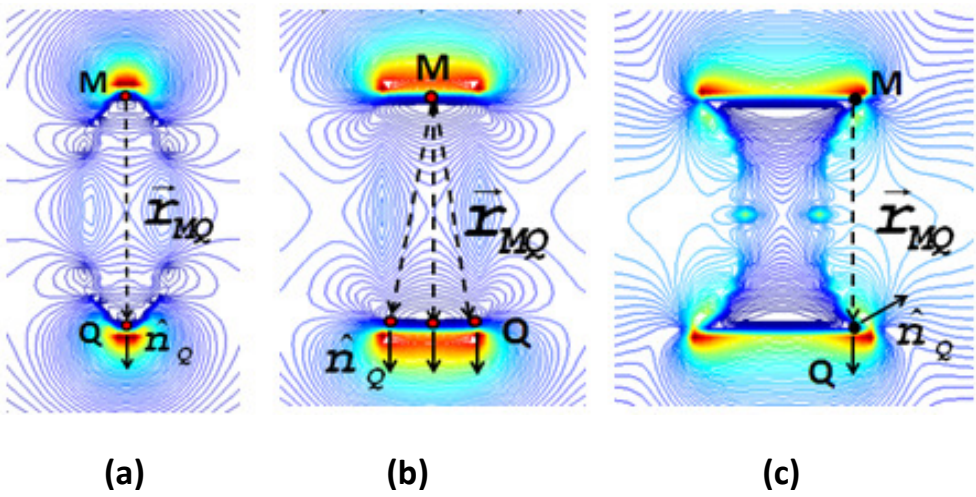

(a)            (b)            (c)